\documentclass[preprintnumbers, prd, showpacs, floatfix, preprintnumbers, letterpaper, nofootinbib, amsmath, amssymb, superscriptaddress,preprint]
{revtex4-1}
\usepackage{graphicx}
\usepackage{epsfig}
\usepackage{bm}
\usepackage{amsfonts}

\usepackage{color}

\newbox\pippobox

\newcommand{\rotating}{}

\def\be{\begin{equation}}
\def\ee{\end{equation}}
\def\ba{\begin{eqnarray} }
\def\ea{\end{eqnarray}}

\newcommand {\lla} {\ {\raise-.5ex\hbox{$\buildrel<\over\sim$}}\ }

\usepackage[T1]{fontenc}
\usepackage[latin1]{inputenc}
\usepackage{graphicx}
\usepackage[english]{babel}
\usepackage{amsmath}
\usepackage{amssymb}
\usepackage{amsfonts}

\def\be{\begin{equation}}
\def\ee{\end{equation}}
\def\bea{\begin{eqnarray}}
\def\eea{\end{eqnarray}}

\begin{document}

\title{Black string in dRGT massive gravity}

\author{Lunchakorn Tannukij \footnote{Email: l\_tannukij@hotmail.com}}
\affiliation{ Department of Physics, Faculty of Science,  Mahidol University, Bangkok 10400, Thailand}
\affiliation{The institute for fundamental study, Naresuan University, Phitsanulok 65000, Thailand}

\author{Pitayuth Wongjun \footnote{Email: pitbaa@gmail.com}}
\affiliation{The institute for fundamental study, Naresuan University, Phitsanulok 65000, Thailand}
\affiliation{Thailand Center of Excellence in Physics, Ministry of Education,
Bangkok 10400, Thailand}

\author{Sushant G. Ghosh \footnote{Email: sghosh2@jmi.ac.in, sgghosh@gmail.com }}
\affiliation{Centre of Theoretical Physics and Multidisciplinary Centre for Advanced Research and Studies (MCARS), Jamia Millia Islamia, New Delhi 110025, India}
\affiliation{Astrophysics and Cosmology Research Unit, School of Mathematical Sciences, University of Kwazulu-Natal, Private Bag 54001, Durban 4000, South Africa}


\begin{abstract}
We present a cylindrically symmetric solution, both charged and uncharged, which is known as a black string solution to the nonlinear ghost-free massive gravity found by de Rham, Gabadadze, and Tolley (dRGT). This ``dRGT black string'' can be thought of as a generalization of the black string solution found by Lemos \cite{1}. Moreover, the dRGT black string solution also include other classes of black string solution such as the monopole-black string ones since the graviton mass contributes to the global monopole term as well as the cosmological constant term. To investigate the solution, we compute mass, temperature, and entropy of the dRGT black string. We found that the existence of the graviton mass drastically affects the thermodynamics of the black string. Furthermore, the Hawking-Page phase transition is found to be possible for the dRGT black string as well as the charged dRGT black string. In terms of their stability, the dRGT black string solution is thermodynamically stable for $r>r_c$ with negative thermodynamical potential and positive heat capacity while it is unstable for $r<r_c$ where the potential is positive.
\end{abstract}

\maketitle

\section{Introduction}
In general relativity (GR),  where the corresponding graviton is a massless spin-2 particle, is the current description of gravitation in physics
and has importantly astrophysical implications. One of the alternating theory of gravity, which follows a very simple idea of generalization to GR, is known as a massive gravity where interaction terms, corresponding to graviton mass, are added into GR. The result of such an introduction is that the gravity is modified significantly at large scale so that, in the cosmological point of view, a system like our universe can enter the cosmic accelerating expansion phase. Not only that, the gravity at smaller scale is not altered much by such modification to ensure the same predictions as GR does.  

However, adding generic mass terms for the graviton on given background usually brings various instabilities and ghosts for the gravitational theories.   To avoid the appearance of the ghost in massive gravity suggested by Boulware and Deser \cite{Boulware:1973my}, the set of possible interaction terms is  constructed accordingly by de Rham, Gabadadze and Tolley (dRGT) \cite{deRham:2010ik,deRham:2010kj}. In particular, the interaction terms are constructed in a specific way to ensure that the corresponding equations of motion are at most second order differential equations so that there is no ghost field. The allowed interaction terms for the four-dimensional theory consist of three kinds of combination; the quadratic, the cubic, and the quartic orders of the metric. Unfortunately, such nonlinearities involve complexities in calculation as a price of eliminating the ghost and usually make it cumbersome to find an exact solution. Nevertheless, there have been plenty interesting approaches to tackle with the complexity to obtain spherical symmetric solutions in massive gravity theories \cite{ebcd,asjs,Ghosh:2015cva,Adams:2014vza, Xu:2015rfa,ccnp,msv12, bcnp,tmn, Berezhiani:2011mt,Cai:2012db,ebaf,Babichev:2015xha,Capela:2011mh, bcp,msv13,tkn,Vegh:2013sk,Cai:2014znn, Koyama:2011yg,Koyama:2011xz,Sbisa:2012zk}. In particular, it was found by Vegh \cite{Vegh:2013sk} that a spherically symmetric black hole with a Ricci flat horizon is a solution to the four-dimensional massive gravity and then the solution was studied in the aspect of thermodynamical properties and phase transition structure \cite{Cai:2014znn,Ghosh:2015cva}.  The spherically symmetric  solutions for dRGT were also considered in \cite{tmn,bcp}. The extension of the solution in terms of electric charge was found in \cite{Berezhiani:2011mt,Cai:2012db}. Moreover, the bi-gravity extension of the solution was found in \cite{ebaf} which covers the previously known spherically symmetric solutions (See \cite{msv13,tkn,Babichev:2015xha}, for reviews on black holes in massive gravity). Recently, other extensions of dRGT massive gravity solution are explored \cite{Hu:2016hpm, Hu:2016mym, Zou:2016sab, Hendi:2017arn, Hendi:2017ibm, Hendi:2016usw, EslamPanah:2016pgc, Hendi:2016hbe, Hendi:2016uni, Hendi:2016yof}.

In astrophysics, it is commonly known that  black holes can be created by the gravitational collapse of massive stars. However, the famous Thorne's \cite{6} hoop conjecture  states  that gravitational collapse of a massive star will yield a black hole only if the mass $ M $ is compressed to a region with circumference $ C <4 \pi M $ in all directions. If the hoop conjecture is true, cylindrical matter will not form a black hole. However, the hoop conjecture was given for spacetime with a zero cosmological constant.  When a negative cosmological constant is introduced, the spacetime will become asymptotically anti-de Sitter spacetime. Indeed, Lemos  \cite{7}  has shown a possibility of cylindrical black holes  from the gravitational collapse with cylindrical matter distribution in an anti-de Sitter spacetime, violating in this way the hoop conjecture.  This cylindrically static black hole solutions in  an anti-de Sitter spacetime are called as black strings.   The pioneering works  on  black string are due to Lemos \cite{1,Lemos:1994xp,3}, and soon  its charged  and rotating \cite{4} counterpart were also obtained.

The main purpose of this paper is to present an exact black string solution including its generalization to charge case in the dRGT massive gravity, and also to discuss their thermodynamical properties with a focus on thermodynamic stability. The possibility to obtain the Hawking-Page phase transition is explored.
We use units which fix the speed of light and the gravitational constant via $G = c = 1$, and use the metric signature ($-,\;+,\;+,\;+$).

\section{\MakeLowercase{d}RGT massive gravity}\label{model}
In this section we review an important ingredient of dRGT massive gravity theory, which is a well known nonlinear generalization of a massive gravity and is free of the Boulware-Deser ghost by incorporating higher order interaction terms into the Lagrangian. The dRGT Massive gravity action is the well-known Einstein-Hilbert action plus suitable nonlinear interaction terms as given by \cite{deRham:2010kj}
\begin{eqnarray}\label{action}
 S = \int d^4x \sqrt{-g}\; \frac{1}{2} \left[ R(g) +m_g^2\,\, {\cal U}(g, f)\right],
\end{eqnarray}
where $R$ is the Ricci scalar  and ${\cal U}$ is a potential for the graviton which modifies the
gravitational sector with the parameter $m_g$ interpreted as  graviton mass.
The effective potential ${\cal U}$ in four-dimensional spacetime is given by
\begin{eqnarray}\label{potential}
 {\cal U}(g, \phi^a) = {\cal U}_2 + \alpha_3{\cal U}_3 +\alpha_4{\cal U}_4 ,
\end{eqnarray}
where $\alpha_3$ and $\alpha_4$ are dimensionless free parameters of the theory.
The terms ${\cal U}_2$, ${\cal U}_3$ and ${\cal U}_4$  can be written in terms of the physical metric $g_{\mu\nu}$ and the reference (or fiducial) metric  $f_{\mu\nu}$ as
\begin{eqnarray}
 {\cal U}_2&\equiv&[{\cal K}]^2-[{\cal K}^2] ,\\
 {\cal U}_3&\equiv&[{\cal K}]^3-3[{\cal K}][{\cal K}^2]+2[{\cal K}^3] ,\\
 {\cal U}_4&\equiv&[{\cal K}]^4-6[{\cal K}]^2[{\cal K}^2]+8[{\cal K}][{\cal
K}^3]+3[{\cal K}^2]^2-6[{\cal K}^4],
\end{eqnarray}
where
\begin{eqnarray}
 {\cal K}^\mu_\nu =
\delta^\mu_\nu-\sqrt{g^{\mu\nu}f_{\mu\nu}}, \label{K-tensor}
\end{eqnarray}
where the rectangular brackets denote the traces,
namely $[{\cal K}]={\cal K}^\mu_\mu$ and $[{\cal K}^n]=({\cal K}^n)^\mu_\mu$. Note that we choose the unitary gauge in our consideration \cite{Vegh:2013sk}. One can see that there are no kinetic terms for the fiducial metric. It seems to play the role of the Lagrange multiplier to help us to construct the suitable form of the mass term.  Conveniently, we redefine the two
parameters $\alpha_3$ and $\alpha_4$ of the graviton potential in Eq. \eqref{potential} by introducing two new parameters $\alpha$ and
$\beta$, as follows,
\begin{eqnarray}\label{alphabeta}
 \alpha_3 = \frac{\alpha-1}{3}~,~~\alpha_4 =
\frac{\beta}{4}+\frac{1-\alpha}{12}.
\end{eqnarray}

The equation of motion of the theory can be obtained by varying the action with respect to dynamical metric $g_{\mu\nu}$ as
\begin{eqnarray}\label{EoM}
 G_{\mu\nu} +m_g^2 X_{\mu\nu} = 0, \label{modEFE}
\end{eqnarray}
where $X_{\mu\nu}$ is  obtained by varying the potential term with respect to $g_{\mu\nu}$ known as the effective energy-momentum tensor. This effective energy-momentum tensor can be written explicitly as 
\begin{eqnarray}
 X_{\mu\nu} &=& {\cal K}_ {\mu\nu} -{\cal K}g_ {\mu\nu} -\alpha\left({\cal K}^2_{\mu\nu}-{\cal K}{\cal K}_{\mu\nu} +\frac{{\cal U}_2}{2}g_{\mu\nu}\right) +3\beta\left( {\cal K}^3_{\mu\nu} -{\cal K}{\cal K}^2_{\mu\nu} +\frac{{\cal U}_2}{2}{\cal K}_{\mu\nu} - \frac{{\cal U}_3}{6}g_{\mu\nu} \right), \,\,\,\,\,\, \label{effemt}
\end{eqnarray}
and obeys {a} constraint by using the Bianchi identities as follows,
\begin{eqnarray}\label{BiEoM}
 \nabla^\mu X_{\mu\nu} = 0,
\end{eqnarray}
where $\nabla^\mu$ denotes the covariant derivative which is compatible with $g_{\mu\nu}$. In the next section, we will find the solution to this equation of motion by imposing the static and axial symmetry.

\section{Black string solution in \MakeLowercase{d}RGT massive gravity} \label{solutions}
In this section, we will look for a static and axially symmetric solution of the modified Einstein equations  (\ref{EoM})  with the following physical metric ansatz,
\begin{eqnarray}\label{metric-gen}
 ds^2 = -n(r)dt^2 + 2d(r) dt dr +\frac{dr^2}{f(r)} + L(r)^2 d\Omega^2,
\end{eqnarray}
and with the Minkowski flat fiducial metric,
\begin{eqnarray}
ds^2 = -dt^2 + dr^2 + r^2 d\Omega^2.
\end{eqnarray}
Here $d\Omega^2 =d\varphi^2 + \alpha_g^2 dz^2$ is a metric on the 2-D surface. The 2-D surface is chosen to be compatible with black string solution \cite{Lemos:1994xp,Cai:1996eg}. The solutions in this ansatz can be classified into two branches: $d(r) = 0$ or $L(r) = l_0 r$ where $l_0$ is a constant in terms of the parameters $\alpha$ and $\beta$ as
\begin{eqnarray}
l_0 = \frac{ \left(\alpha +3 \beta \pm \sqrt{\alpha ^2-3 \beta } \right)}{2 \alpha +3 \beta +1}.
\end{eqnarray}
From this stage, it is convenient to investigate in the branch $d = 0$. Even though we choose to investigate the simple branch, the equation is still difficult to investigate analytically. Since the fiducial metric seems to play the role of a Lagrange multiplier to eliminate the BD ghost, one can choose an appropriate form to simplify the calculation. Indeed, the suitable form of the fiducial metric significantly provides the analytical form of the physical metric, for example \cite{Vegh:2013sk} for spherically symmetric solution and \cite{Chullaphan:2015ija} for cosmological solution.
In the present work, we choose the fiducial metric  {to be}
\begin{eqnarray}\label{fiducial metric}
f_{\mu\nu}=\text{diag}(0,0,h(r)^2  ,\alpha_f^2 h(r)^2 ),
\end{eqnarray}
where $h(r)$ is an arbitrary function.  The line element of the corresponding physical metric in this branch can be written as
\begin{eqnarray}\label{metric}
 ds^2 = -n(r)dt^2 +\frac{dr^2}{f(r)} +L(r)^2 d\Omega^2,
\end{eqnarray}
From this ansatz, components of the Einstein tensor can be written as
\begin{eqnarray}
G^{t}_{t} &=& \frac{L f' L'+2 f L L''+f L'^2}{L^2},\\
G^{r}_{r} &=& \frac{f L' \left(n L'+L n'\right)}{L^2 n},\\
G^{\varphi}_{\varphi} = G^{z}_{z} &=& \frac{n f' \left(2 n L'+L n'\right)+f \left(4 n^2 L''+2 n L' n'+2 L n n''-L n'^2\right)}{4 L n^2}. \label{EFE}
\end{eqnarray}
Computing the effective energy-momentum tensor in Eq. (\ref{effemt}) with this ansatz, the tensor $X_{\mu\nu}$ can be written as
\begin{eqnarray}
X^{t}_{t} = X^{r}_{r} &=& \frac{h \alpha_f}{L \alpha_g }-\frac{3 L-h}{L}+ \alpha  \left(\frac{h \alpha_f (2 L-h)}{L^2 \alpha_g }-\frac{3 L-2 h}{L}\right)\nonumber \\
&+&\beta  \left(\frac{3 h \alpha_f (L-h)}{L^2 \alpha_g }-\frac{3 (L-h)}{L}\right),\\
X^{\varphi}_{\varphi} &=& \frac{h \alpha_f}{L \alpha_g }-3 +\alpha  \left(\frac{2 h \alpha_f}{L \alpha_g }-3\right)+\beta  \left(\frac{3 h \alpha_f}{L \alpha_g }-3\right),\label{effemt2}\\
X^{z}_{z} &=&-\left(\frac{h-3 L}{L}+\frac{\alpha  (2 h-3 L)}{L}+\frac{3 \beta  (h-L)}{L}\right). \label{effemt3}
\end{eqnarray}

\begin{figure}[h!]
\begin{center}
\includegraphics[scale=0.8]{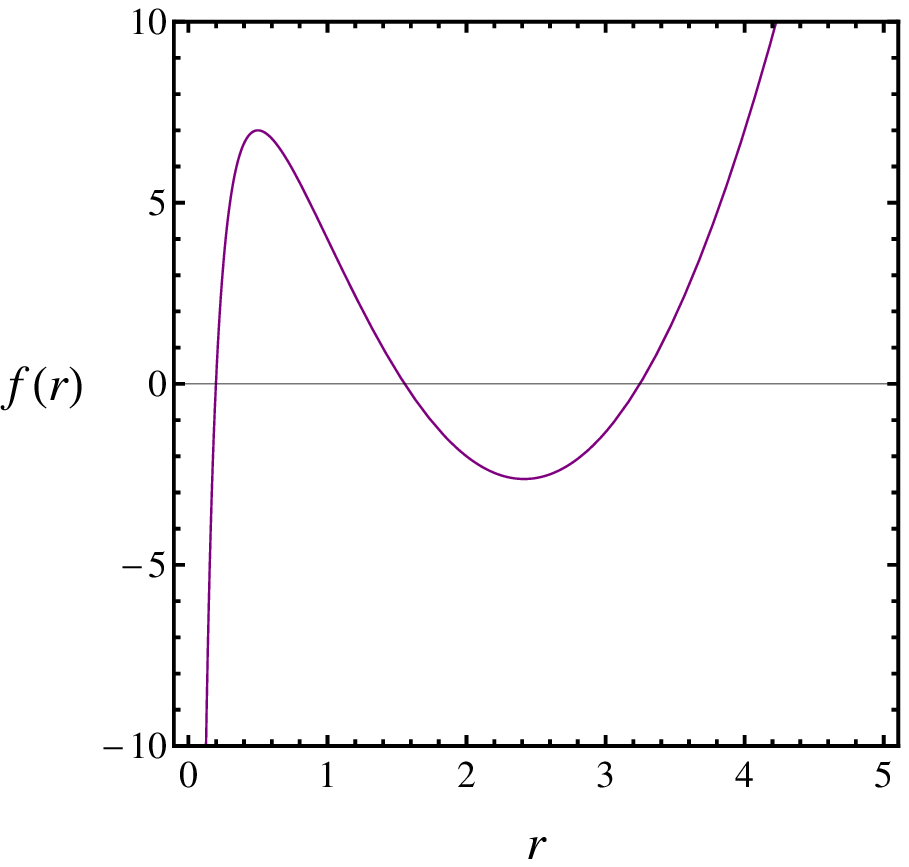}\qquad
\includegraphics[scale=0.8]{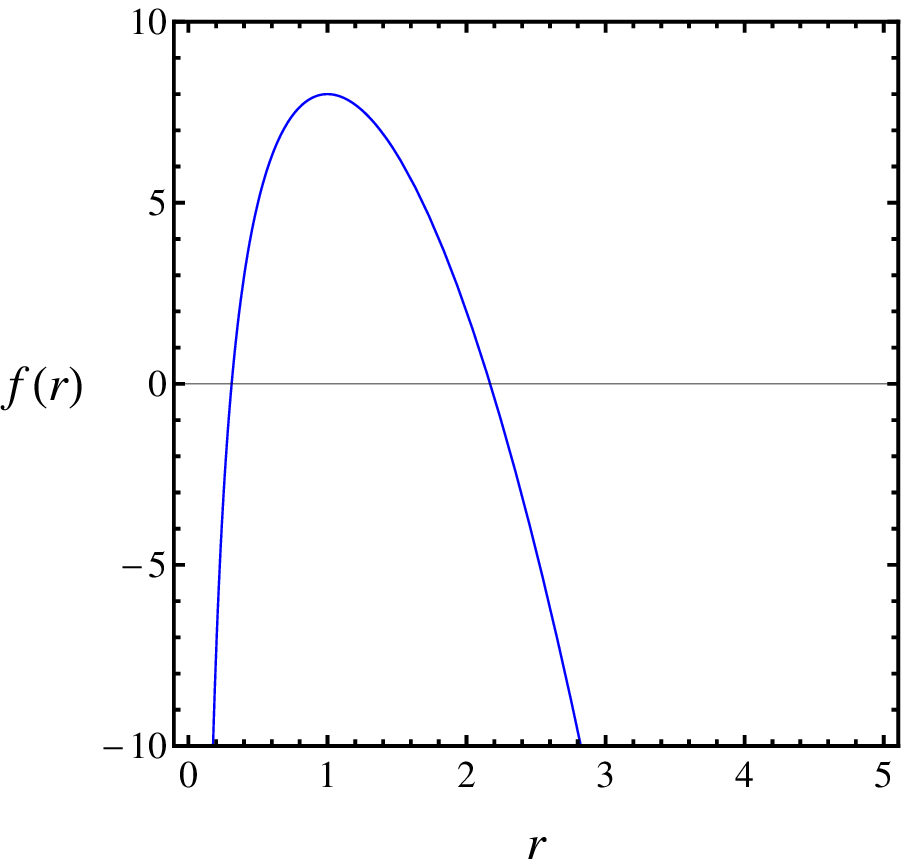}
\end{center}
{\caption{The left panel shows the horizon structure of the AdS solution for specific values of the parameters being $\alpha_m^2 = 4, c_0 = 6, c_1=5$. The right panel shows the horizon structure of the dS solution for the value of of the parameters as $\alpha_m^2 = -4, c_0 = -3, c_1=1$.}\label{fig:dSAdShorizon}}
\end{figure}

By substituting these components into modified equation in Eq. (\ref{modEFE}) and then using the equations in $(t,t)$ and $(r,r)$ components, we found the constraint such that
\begin{eqnarray}
\frac{d}{dr}\left(\frac{f L'^2}{n}\right) =0.
\end{eqnarray}
In order to obtain the black hole solution with $f =n$, the function $L(r)$ must be proportional to $r$. Therefore, we can set $L =r$ for the following investigation.  By using this setting, we have only two independent functions, $f$ and $h$. The two independent equations for this two functions are the conservation of the energy momentum tensor in Eq. (\ref{BiEoM}) and the $(t,t)$ component of the modified Einstein equation in Eq. (\ref{modEFE}) which can be expressed respectively as
\begin{gather}
\frac{h' (r (1+2 \alpha +3 \beta ) (\alpha_f+\alpha_g )-2 h \alpha_f(\alpha +3 \beta ))}{\alpha_g  r^2}= 0 ,\label{eom1}
\\
\frac{r f'+f}{r^2}= -m_{g}^2 \Bigg(\frac{h \alpha_f}{\alpha_g  r}-\frac{3 r-h}{r}+\alpha  \left(\frac{h \alpha_f (2 r-h)}{\alpha_g  r^2}-\frac{3 r-2 h}{r}\right)\nonumber \\
+ \beta  \left(\frac{3 h \alpha_f (r-h)}{\alpha_g  r^2}-\frac{3 (r-h)}{r}\right)\Bigg),\label{eom2}
\end{gather}
From Eq. (\ref{eom1}), two exact solutions of $h$ can be written as
\begin{eqnarray}
h(r) &=&\frac{ r (1+2 \alpha +3 \beta )(\alpha_f+\alpha_g )}{2 \alpha_f (\alpha +3 \beta )}, \,\text{and}\label{solutionh1}\\
h(r) &=& h_0 = \text{constant},\label{solutionh2}
\end{eqnarray}
with these solutions of $h(r)$, Eq. ~(\ref{eom2}) admits the following two solutions
\begin{eqnarray}
f_1(r) &=&\left(-\frac{m_{g}^2  \left(1+\alpha +\alpha ^2-3 \beta \right)}{3 (\alpha +3 \beta) }\right)r^2-\frac{b}{\alpha_g r}, \label{solutionf1}\\
f_2(r) &=&\left(m_{g}^2 (1+\alpha +\beta )\right)r^2-\frac{b}{\alpha_g r} -m_{g}^2 h_0 (1+2 \alpha +3 \beta )r +m_{g}^2 h_0^2 (\alpha +3 \beta ) , \label{solutionf2}
\end{eqnarray}
where $b$ is an integration constant. Note that we have used the component $(\varphi,\varphi)$ of the modified Einstein equation in Eq. (\ref{modEFE}) to obtain the constraint $\alpha_f = \alpha_g$. The first solution in Eq. ~ (\ref{solutionf1}) coincides with the black string solution in general relativity \cite{1,Lemos:1994xp,Cai:1996eg} with an effective cosmological constant $\Lambda$ as
\begin{eqnarray}
\Lambda = -\frac{m_{g}^2  \left(1+\alpha +\alpha ^2-3 \beta \right)}{3 (\alpha +3 \beta) }.
\end{eqnarray}
Note that in \cite{1,Lemos:1994xp,Cai:1996eg}, the constant $\alpha_g^2$ is chosen to be $\alpha_g^2 = - \Lambda/3$, and
  \begin{eqnarray}
\alpha_g^2 = \frac{m_{g}^2  \left(1+\alpha +\alpha ^2-3 \beta \right)}{ (\alpha +3 \beta) }.
\end{eqnarray}
The integration constant $b$ can be obtained by using the solutions in Newtonian limit which can be expressed as $b = 4M$, where $M$ is the ADM mass per unit length in $z$ direction.
Mathematically, the solution (\ref{solutionf1}) is exactly the same as one in general relativity which is already widely investigated.
Moreover, it's charged and rotating counterparts, and  their properties including  thermodynamics have been also widely investigated \cite{1,Lemos:1994xp,3,4,6,7}.
Although the solution (\ref{solutionf1}) coincides mathematically with that of Lemos \cite{1,Lemos:1994xp,Cai:1996eg},  the dRGT solution naturally generates  the cosmological-constant-like term from graviton mass while for Lemos' \cite{1} black string solution the cosmological constant is introduced by hand. Hence, the properties of solution (\ref{solutionf1}) can be analyzed as in the previously mentioned cases, and hence we shall not do them here.   The second solution in Eq. (\ref{solutionf2}) is new. We can redefine the variables and parameters as follows,
\begin{eqnarray}
f_2({r}) &=&\alpha_{m}^2r^2-\frac{4M}{\alpha_g r}-\alpha_{m}^2c_1 r +\alpha_{m}^2c_0,\label{solutionf2-new}
\end{eqnarray}
where
\begin{eqnarray}
\alpha_{m}^2 \equiv m^2_g \left(1+\alpha +\beta \right),\qquad c_1 \equiv \frac{h_0  (1+2 \alpha +3 \beta )}{1+\alpha +\beta}, \qquad c_0 \equiv \frac{h_0^2 (\alpha +3 \beta )}{1+\alpha +\beta}. \label{defineparameter}
\end{eqnarray}
The solution in Eq. (\ref{solutionf2-new}) is an exact black string solution in dRGT massive gravity which, in the limit $\alpha_m = \alpha_g$ and $c_0=c_1=0$, naturally goes over to Lemos' \cite{1} black string in general relativity. Henceforth, for definiteness, we shall call this massive gravity solution (\ref{solutionf2-new})  as  dRGT black string.  In particular, it incorporates the cosmological constant term  naturally in terms of the graviton mass $m_g$ which should not be surprising since the graviton mass serves as the cosmological constant in the self-expanding cosmological solution in massive gravity. The graviton mass also generates a constant term which is known as the so-called global monopole term.
One can see that the horizon structure depend on the sign of $\alpha_m^2$. If $\alpha_m^2>0$, corresponding to Anti de Sitter-like solution, the maximum number of the horizons are three. If $\alpha_m^2<0$, corresponding to de Sitter-like solution, the maximum number of the horizons are two. The generic behavior of the horizon structure are shown in Fig. \ref{fig:dSAdShorizon}.  This solution modifies the black string solution in general relativity with cosmological constant \cite{1}  by three parameters $c_0, c_1$ and $\alpha_m^2$.

\subsection{Thermodynamics of dRGT black string}
Now, we analyze the thermodynamics of  the new dRGT black string (\ref{solutionf2-new}). By definition,  the  horizons are zeros of $f(r_+)=0$. Depending on the parameters, there may exist upto three real roots.
The horizon radius ($r_+$) and black string mass ($M_+$) are related via
\begin{eqnarray}
M_+ = \frac{1}{4} \alpha_g \alpha^2_m r_+ \left(r^2_+ -c_1 r_+ + c_0\right).
\end{eqnarray}
The mass $M_+$ depends on parameters $c_0,\; c_1,\; \alpha^2_m$ and $\alpha^2_g$. To have a positive definite mass,
one may require the term $\alpha^2_m r_+ \left(r^2_+ -c_1 r_+ + c_0\right)$ being positive. Such a consideration yields the following conditions,
\begin{eqnarray}
\alpha^2_m > 0, \qquad c_1^2-4c_0<0. \label{Mpositivecond}
\end{eqnarray}
The second condition is obtained from a requirement that $r^2_+ -c_1 r_+ + c_0$ cannot be factorized so that, including the condition $\alpha^2_m > 0$, the term $\alpha^2_m r_+ \left(r^2_+ -c_1 r_+ + c_0\right)$ will be positive definite.
The Hawking temperature of the dRGT black string can be obtained by using $T = \kappa/(2\pi)$, where $\kappa = f'(r)/2|_{r=r_+}$ is the surface gravity of the black string. As a result, the temperature can be written as
\begin{eqnarray}
T_+=\frac{\alpha_m^2}{4\pi r_+}\left(3r_+^2-2c_1 r_+ +c_0\right).
\end{eqnarray}
We can determine a condition for a positive definite temperature by  using the same procedure as we did to the mass. Consequently, we obtain
\begin{eqnarray}
c_1^2-3c_0<0. \label{Tpositivecond}
\end{eqnarray}
These positivity conditions for both the mass and the temperature can narrow the parameter space down. The minimum temperature can be found at $r_+ = \sqrt{c_0/3}$. At this point, the temperature can be written as
\begin{align}
T_{min} = \frac{\alpha_m^2}{2\pi}\left(\sqrt{3c_0}-c_1\right).
\end{align}
The square root of $c_0$ is well-defined through the conditions in Eq. (\ref{Mpositivecond}) and/or in Eq. (\ref{Tpositivecond}).
Next, we can find its entropy via the first law of thermodynamics $dM_+=T_+dS$ which reads
\begin{align}
S &= \frac{1}{2}\pi \alpha_g r_+^2 = \frac{A}{4},
\\
A &= 2\pi \alpha_g r_+^2,
\end{align}
and it turns out to satisfy the area law.
Now we find the thermodynamical stability of the black string. For the locally thermodynamical stability, we find the heat capacity of the system which can be expressed as
\begin{align}
C = \left(T \frac{dS}{dT}\right)_{r=r_+} = \frac{\pi \alpha_g r_+^2\left(3r_+^2 -2c_1 r_+ +c_0\right)}{3r_+^2-c_0}.
\end{align}
By the existence of the minimum temperature, $C$ diverges at that temperature, where the corresponding radius obeys $r_+ = \sqrt{c_0/3}$. Moreover, the heat capacity is negative when $r_+ < \sqrt{c_0/3}$ and positive when $r_+ > \sqrt{c_0/3}$. This suggests that the locally thermodynamical stability of the black string requires the condition $r_+ > \sqrt{c_0/3}$. Apart from the locally one, the globally thermodynamical stability can be analyzed by considering the Helmholtz potential
\begin{align}
F &= M-TS,
\\
&= \frac{\alpha_g \alpha_m^2 r_+}{8}\left( c_0-r_+^2\right).
\end{align}
The black string will be stable if the Helmholtz potential is negative. Thus, the globally thermodynamical stability condition can be expressed as \begin{align}
r_+ > r_c = \sqrt{c_0}.
\end{align}
Note that this condition also satisfies the local stability condition. As a result, the Hawking-Page phase transition, the transition from the hot flat space state to the black hole or black string state \cite{Hawking:1982dh}, occurs at the critical point, $r_+=r_c=\sqrt{c_0}$. It is interesting to note that the phase transition can be realized only for the asymptotic AdS solution. Moreover, the existence of the phase transition depends only on the graviton mass. This can be seen clearly when consider the case $c_1 = c_0 = 0$. In this case, Helmholtz potential does not change sign which means that the Lemos black string is always thermodynamically stable.

\section{Charged \MakeLowercase{d}RGT black string}
Even though the astronomical objects are neutral, the charged black holes are intensively investigated. For spherically symmetric solutions, the charged black hole is a useful toy model since the structure of the horizon
usually differs that of the neutral one in a quite drastic way.
Moreover, a study of thermodynamics of the charged black holes can be significantly extended since it usually involve the open system such as grand canonical system where the charge can play the role of the chemical potential. In this section, we extend our consideration by adding the electric charge into the theory. As a result, the action can be obtained by adding the a gauge field term as follows,
\begin{eqnarray}
S = \int d^4x \sqrt{-g}  \Bigg[\frac{1}{2}\left[ R +m_g^2 {\cal U}(g, \phi^a)\right]-\frac{1}{16\pi}F_{\mu\nu}F^{\mu\nu}\Bigg], \label{actioncharge}
\end{eqnarray}
where $F_{\mu\nu}\equiv\left(\nabla_\mu A_\nu-\nabla_\nu A_\mu\right)$,  $A_\mu=(a(r),0,0,0)$ and $a(r)$ is an arbitrary function. By using the same procedure as in non-charged case, the solutions still can be divided into two branches according to Eq. (\ref{solutionh1}) and Eq. (\ref{solutionh2}). The equations of motion of the electric field, which is the Maxwell equations, provides the constraint to the function $a(r)$ as
\begin{eqnarray}
a(r) =- \frac{ \gamma }{\alpha_g r},
\end{eqnarray}
where $\gamma$ is an integration constant. In order to reduce the theory to the electromagnetic theory in flat spacetime, the integration constant can be written in terms of the linear charge density, $q$, in $z$ direction as $\gamma^2 = 4 q^2$. By solving the modified Einstein equations, the solutions for both branches can be written as
\begin{eqnarray}
f^{(q)}_1(r) &=&\left(-\frac{m_{g}^2  \left(1+\alpha +\alpha ^2-3 \beta \right)}{3 (\alpha +3 \beta) }\right)r^2-\frac{b}{\alpha_g r}+\frac{\gamma^2}{\alpha_g^2 r^2} , \label{solutionf-c1}\\
f^{(q)}_2(r) &=& \left(m_{g}^2 (1+\alpha +\beta )\right)r^2-\frac{b}{\alpha_g r}+\frac{\gamma^2}{\alpha_g^2 r^2} -m_{g}^2 h_0  (1+2 \alpha +3 \beta )r+m_{g}^2 h_0^2 (\alpha +3 \beta ). \label{solutionf-c2}
\end{eqnarray}
As we have mentioned before, we will focus on the second branch of the solutions which can be written in terms of the parameters in Eq. (\ref{defineparameter}) as
\begin{eqnarray}
f(r) &=&\alpha_m^2 r^2-\frac{4 M}{\alpha_g r}+\frac{4 q^2}{\alpha_g^2 r^2}- \alpha_m^2 c_1 r + \alpha_m^2  c_0.\label{solutionfC}
\end{eqnarray}
The solution (\ref{solutionfC}) is the charged dRGT black string solution.
The linear mass density $M$ can be expressed by solving $f(r_+) = 0$ as follows,
\begin{eqnarray}
M^{(q)} = \frac{1}{4} \alpha_g \alpha_m^2 r_+ \left(r_+^2-c_1 r_+ + c_0\right) + \frac{q^2}{\alpha_g r_+}.
\end{eqnarray}
For a charged dRGT black string, it can be treated as  two kinds of thermodynamical system; grand canonical and canonical systems. We will separate our considerations into two following subsections.

\subsection{Grand Canonical black string}
In this aspect of treatment, a charged dRGT black string is considered to be an open system where charge transfer or creation/annihilation are allowed. The energy gain/loss from the increase/decrease of charge is determined by the chemical potential,
\begin{align}
\mu = \frac{2 q}{\alpha_g r},
\end{align}
evaluating at the horizon. The charged dRGT black string mass and the Hawking temperature can be written as
\begin{eqnarray}
M^{(q)}_G &=& \frac{1}{4} \alpha_g \alpha_m^2 r_+ \left(r_+^2-c_1 r_+ + c_0\right) +\frac{1}{4}\alpha_g r_+ \mu^2  ,\\
T^{(q)}_G &=& \frac{\alpha_m^2\left(3r^2_+-2c_1 r_+ +c_0\right) -\mu^2}{4\pi r_+}.
\end{eqnarray}
The positive definite of both mass and temperature can be found by following the same step as done in the non-charge case. As a result, the condition can be expressed as
\begin{eqnarray}
\alpha_m^2 >0, \quad c_1^2 - 3 \left(c_0 -\frac{\mu^2}{\alpha_m^2}\right) < 0.
\end{eqnarray}
From these thermodynamic quantities, one found that they satisfy the first law of thermodynamics and the area law in which the entropy of the charged dRGT black string in the grand canonical system can be written as
\begin{align}
S^{(q)}_G &= \frac{1}{2}\pi \alpha_g r_+^2 = \frac{A}{4},
\end{align}
For the thermodynamical stability of the system, one can analyze by finding the heat capacity and the Gibbs potential. By following the same strategy as done in the non-charge case, the heat capacity  and the Gibbs potential can be found as
\begin{align}
C^{(q)}_G &= \left(T \frac{dS}{dT}\right)_{r=r_+}= \frac{\pi \alpha_g r^2_+\left[\alpha_m^2\left(3r^2_+-2c_1 r_+ +c_0\right) -\mu^2\right]}{\alpha_m^2\left(3r^2_+-c_0\right)+\mu^2},\\
G^{(q)} &= M-TS-\mu Q=-\frac{\alpha_g r_+}{8}\left[\alpha_m^2\left(r^2_+ -c_0\right)+\mu^2\right].
\end{align}
The thermodynamical stability conditions can be found by requiring $C^{(q)}_G > 0$ and $ G^{(q)} < 0$. The resulting condition can be expressed as
\begin{eqnarray}
r_+ > r_c = \sqrt{ \left(c_0 -\frac{\mu^2}{\alpha_m^2}\right) }.
\end{eqnarray}
This suggests that it is possible to obtain the Hawking-Page phase transition for charged dRGT black string in massive gravity theory , when considered in the grand canonical aspect. Note that this phase transition cannot be provided by using Lemos black string solution. This issue can be seen by setting $c_1=c_0=0$ which turns out that the Gibbs potential is always negative for AdS solution while the dS solution does not satisfy the requirement of positive definite of the charged  dRGT black string mass and temperature.

\subsection{Canonical black string}
For the canonical aspect of the black string, charge transfer or charge creation/annihilation are not allowed. The Hawking temperature can be found as
\begin{align}
T^{(q)}_C = \frac{1}{4\pi r_+}\left[\alpha_m^2\left(3r^2_+ -2c_1 r_+ +c_0\right) - \frac{4q^2}{\alpha_g^2 r^2_+}\right].
\end{align}
From this expression, we found that one of requirements to obtain the positive definite of the black string temperature is $\alpha_m^2 >0$. This corresponds to the asymptotic AdS solution. By following the strategy as done before,  we found that the system still obey the first law of thermodynamics and the area law in which the entropy of the black string can be written as
\begin{align}
S^{(q)}_C &= \frac{1}{2}\pi \alpha_g r^2_+ = \frac{A}{4}.
\end{align}
For the stability issue of this thermodynamic system, the local stability can be analyzed by considering the heat capacity  which can be found as
\begin{align}
C^{(q)}_C = \frac{\pi \alpha_g r^2_+\left[\alpha_m^2\left(3r^2_+-2c_1 r_+  +c_0\right)-\frac{4q^2}{\alpha_g^2 r^2_+}\right]}{\alpha_m^2\left(3r^2_+ -c_0\right)+\frac{12q^2}{\alpha_g^2 r^2_+}}.
\end{align}
By using $T^{(q)}_C >0$, the positivity of the heat capacity will be guaranteed by requiring
\begin{eqnarray}
r_+ > \sqrt{ \frac{c_0}{6}\left(1 + \left[1 - \frac{144 q^2}{\alpha_g^2\alpha_m^2 c_0^2}\right]^{1/2}\right) }.
\end{eqnarray}
The Helmholtz potential of the system can be written as
\begin{align}
F^{(q)} = -\frac{\alpha_g r_+}{8}\left[\alpha_m^2\left(r^2_+ -c_0\right)-\frac{12q^2}{\alpha_g^2 r^2_+}\right].
\end{align}
To investigate the globally thermodynamical stability, one requires that $F^{(q)} < 0$. This turns out the condition as follows,
\begin{align}
r_+ > r_c = \sqrt{ c_0\left(1 + \left[1 + \frac{144 q^2}{\alpha_g^2\alpha_m^2 c_0^2}\right]^{1/2}\right) }
\end{align}
which is stronger than one for the local stability condition. This suggests that, for the canonical system, it is possible to provide the Hawking-Page phase transition for the charged black string in massive gravity theory. However, for this system, the AdS black string solution can also provide the phase transition. This can be seen by setting $c_1 = c_0 = 0$ which turns out that the phase transition occur when $r_+^2 = \sqrt{\frac{2 q}{\sqrt{3\alpha_m^2\alpha_g^2}}}$.

\section{Concluding remarks}\label{summary}
The dRGT massive gravity describes nonlinear interaction terms as a generalization of the Einstein-Hilbert action when graviton is massive.  It is believed that dRGT massive gravity may provide a possible explanation for the accelerating expansion of the universe that does not require any dark energy or cosmological constant.
In this paper, we have presented a class of \rotating black string, both charged and uncharged, in dRGT massive gravity where the effective cosmological constant is negative, and studied the thermodynamics and phase structure of the  solutions in both the grand canonical and canonical ensembles.  The black string obtained is immensely simplified due to the choice of the fiducial metric.  As expected the solution contains the Lemos \cite{1} solutions as a particular case.  Our dRGT black string solution can be identified, e.g.,  as monopole black string of general relativity for suitable choice of the parameters of the theory where the graviton mass in massive gravity naturally generates the cosmological constant and the global monopole term.

We have also analyze their thermodynamical properties of the dRGT  black string solution. 
The thermodynamic quantities  have also been found to contain corrections from graviton mass except for the black string entropy which is unaffected by massive gravity and still obeys area law.  By analyzing the thermodynamical properties of the black string solution, we found that it is possible to obtain the Hawking-Page phase transition in dRGT black string while it is not possible for Lemos solution. The conditions to provide such a transition are explored.  In fact, we determine the phase transition in both charged and non-charged case by analyzing the sign of the potential at $r=r_c$, with the stable (unstable) branch for $r > (<) r_c$.

The results presented here are the generalization of previous discussions of Lemos \cite{1,Lemos:1994xp,3}, on the black string, in a more general setting. The possibility of a further generalization of these results to higher dimensions is an interesting problem for future research. The rotating dRGT black string solution is also interesting since most astronomical objects are rotating. We leave this in investigation for further work.

\begin{acknowledgments}
This project is supported by the ICTP through grant  No. OEA-NET-76.  PW is also supported by the Naresuan University Research Fund through grant No. R2559C235. S.G.G. would like to thank SERB-DST Research Project Grant No. SB/S2/HEP-008/2014  and also like to thank IUCAA, Pune for the hospitality, where a part of this work was done.
\end{acknowledgments}

\end{document}